\documentclass[showpacs,prb,aps,superscriptaddress,twocolumn,floatfix]{revtex4-1}
\usepackage[english]{babel}
\usepackage{amsmath}
\usepackage{amssymb,mathrsfs}
\usepackage{graphicx}

\begin{document}

\title{Polarization hydrodynamics in a one-dimensional polariton
  condensate}

\author{P.-\'E.~Larr\'e}\affiliation{Univ.~Paris Sud, CNRS, Laboratoire de
  Physique Th\'eorique et Mod\`eles Statistiques, UMR8626, F-91405
  Orsay, France}
\author{N.~Pavloff}\affiliation{Univ.~Paris Sud, CNRS, Laboratoire de
  Physique Th\'eorique et Mod\`eles Statistiques, UMR8626, F-91405
 Orsay, France}
\author{A.~M.~Kamchatnov}\affiliation{Institute of Spectroscopy,
  Russian Academy of Sciences, Troitsk, Moscow, 142190, Russia}

\begin{abstract}
  We study the hydrodynamics of a nonresonantly-pumped polariton
  condensate in a quasi-one-dimensional quantum wire taking into
  account the spin degree of freedom. We clarify the relevance of the
  Landau criterion for superfluidity in this dissipative two-component
  system. Two Cherenkov-like critical velocities are identified
  corresponding to the opening of different channels of radiation: one
  of (damped) density fluctuations and another of (weakly damped)
  polarization fluctuations.  We determine the drag force exerted onto
  an external obstacle and propose experimentally measurable
  consequences of the specific features of the fluctuations of
  polarization.
\end{abstract}


\maketitle

The condensation of exciton-polaritons in semiconductor microcavity
\cite{Kas06,Bal07,Lai07,Chr07} arouse a great interest directed
towards the possible demonstration of superfluid dynamics in coupled
light-matter waves.  Beautiful experiments revealed suppression of
back-scattering from an obstacle \cite{Amo09a,Amo09b}, nucleation of
quantized vortices \cite{Lag08,Rou11,Nar11,San11}, and generation of
effectively stable oblique solitons \cite{Amo11,Gro11} (see also the
review article \onlinecite{CC-2013} and references therein). Although
the definition of a genuine superfluid behavior in these systems is
still a matter of active debates (see, e.g., the exchange in
Refs.~\onlinecite{But12-Dev12}), it makes no doubt that the coherent
wave-mechanical flow of an exciton-polariton condensate offers the
prospect of studying a rich variety of remarkable hydrodynamic
effects. Among these, the specific features associated to the spin
of the exciton-polaritons are of particular interest. In the
hydrodynamic context they have been revealed by the observation of the
optical spin Hall effect \cite{Ley07}, of half vortices \cite{Lag09},
and of half solitons \cite{Hiv12}.

In the present work we concentrate on linear spin effects; nonlinear
effects are addressed in an other publication \cite{Kam13}. We
describe the polariton condensate by a two-component order parameter
$\psi_{\pm}$ accounting for the spin degree of freedom, corresponding
to the two possible excitonic spin projections $\pm 1$ onto the
structure growth axis and to the right and left circular polarization
of emitted photons. Interactions within the system can be described by
two constants $\alpha_1$ and $\alpha_2$ corresponding to interactions
between polaritons with parallel ($\alpha_1$) or antiparallel
($\alpha_2$) spins. It is accepted that $\alpha_1>0$ and that
$|\alpha_2| < \alpha_1$ [see the discussion after
Eqs.~\eqref{eq:polar2}].  In the following we always consider the
standard situation where $0< -\alpha_2 < \alpha_1$.  In the presence
of an external magnetic field applied parallel to the structure axis,
there is a Zeeman splitting $2\,\hbar\,\Omega$ between the two
circularly polarized states $\psi_+$ and $\psi_-$ (we neglect the
possible small residual splitting of linear polarization considered
for instance in Ref.~\onlinecite{She07}). Taking into account the
effect of the external magnetic field and of the interactions amongst
polaritons, one can write the energy density of the uniform system as
\cite{Rub06}
\begin{equation}\label{eq:polar1}
E=-\hbar\,\Omega \, (\rho^0_+-\rho^0_-) +
\frac{\alpha_1}{2}\big[(\rho^{0}_+)^2+(\rho^{0}_-)^{2}\big] +
\alpha_2 \, \rho^0_+ \, \rho^0_- ,
\end{equation}
where $\rho^0_\pm=|\psi^0_\pm|^2$ is the (uniform) density of
polaritons with spin $\pm 1$, and in the following we denote the
total density of the polariton gas as $\rho^0 = \rho^0_+ +
\rho^0_-$. Then, minimizing the free energy
of the system, one finds two regimes \cite{Rub06}. For large magnetic
field [$\hbar\, \Omega>\hbar\, \Omega_{\rm crit}=
\frac{1}{2}(\alpha_1-\alpha_2)\rho^0$] the system is circularly
polarized with $\rho_-^0=0$, and the chemical potential reads
$\mu=\alpha_1\rho^0-\hbar\, \Omega$. For lower fields ($\hbar\, \Omega
< \hbar\, \Omega_{\rm crit}$) the po\-la\-ri\-za\-tion gradually
becomes linear when $\Omega$ decreases; in this case
one has
\begin{equation}\label{eq:polar2}
\rho^0_\pm=\tfrac{1}{2}\rho^0 \left(1 \pm \Omega/\Omega_{\rm crit}\right)
\quad\mbox{and}\quad
\mu=\tfrac{1}{2}(\alpha_1+\alpha_2)\,\rho^0 ,
\end{equation}
from which it is clear that, in the absence of magnetic field (that
is, when $\Omega=0$), the system is linearly polarized
\cite{Kas06,Bal07}, a feature that ori\-gi\-nates in the present
phenomenological description from the positiveness of
$\alpha_1-\alpha_2$. The fact that $\alpha_1+\alpha_2>0$ implies that
$\mu>0$, that the uniform polariton gas is stable, and that it
corresponds to an emission blue shift \cite{Kas06,Baj08,Uts08}.

A study of spin dynamics has been done in Ref.~\onlinecite{Fla12} in
the case of a fully polarized ground state. In the present work we
treat instead the weak magnetic field regime \eqref{eq:polar2}, and
study the dynamics of the system in the presence of (i) an external
potential representing an obstacle and/or of (ii) modulations of the
uniform ground state.  We consider a one-dimensional wire-shaped
ca\-vi\-ty structure in which the order parameter is of the form
$\psi_\pm(x,t)$ and we model the dynamics of the system by the
following Gross-Pitaevskii-type equation:
\begin{align}\label{eq:polar3}
i\hbar \, \partial_t \psi_{\pm} = &
-\frac{\hbar^2}{2 m}\, \partial^2_x\psi_{\pm}
+ U_{\rm ext}(x+Vt) \nonumber \, \psi_{\pm}
\mp \hbar\, \Omega\, \psi_\pm
 \nonumber \\
& + (\alpha_1\rho_\pm+\alpha_2\rho_{\mp})\, \psi_\pm +
i\,(\gamma-\Gamma \rho)\, \psi_\pm ,
\end{align}
where $m$ is the polariton effective mass (in the parabolic dispersion
approximation, valid at small momenta) and $\rho_{\pm}(x,t)=
|\psi_{\pm}(x,t)|^2$. $U_{\rm ext}(x+Vt)$ describes an obstacle in
motion at velocity $V$ with respect to the polariton fluid. In
accordance with the description (\ref{eq:polar1}), the effect of the
magnetic field is accounted for in Eq.~\eqref{eq:polar3} by the Zeeman
term $\mp\, \hbar\,\Omega \, \psi_\pm$ and interaction effects are
described by local terms proportional to $\alpha_1$ and $\alpha_2$.
Due to the finite polariton-lifetime, the system needs to be
pumped. Following Refs.~\onlinecite{Wou07,Kee08,Wou08,Wou10} we
schematically describe this effect by the last term of
Eq.~\eqref{eq:polar3}: The term $i\, \gamma\, \psi_\pm$ describes the
combined effects of the incoherent pumping and decay processes;
$\gamma>0$, indicating an overall gain counterbalanced by the term
$-i\,\Gamma \rho\,\psi_\pm$ (where $\Gamma>0$ and
$\rho=\rho_++\rho_-$), which phenomenologically accounts for a
saturation of the gain at large density and makes it possible to reach
a steady-state configuration with a finite density
$\rho^0=\gamma/\Gamma$. Note that the saturation term is proportional
to $\rho$.  Arguing on weak cross-spin scattering, the authors of
Ref.~\onlinecite{Bor10} used a different type of saturation of the
gain, proportional to $\rho_{\pm}$: In this case the
value of the stationary background densities $\rho^0_{+}$ and
$\rho^0_{-}$ is fixed {\it a priori}, independently of the magnetic
field. In the present work we follow Ref.~\onlinecite{Para10} and use
a model where the value of $\rho^0_{+}$ and $\rho^0_{-}$ is fixed by
the thermodynamic equilibrium between the two spin components in the
presence of a magnetic field [Eqs.~\eqref{eq:polar2}].

A small departure from the stationary configuration \eqref{eq:polar2}
is described by an order parameter of the form
\begin{equation}\label{eq:polar4}
\psi_\pm(x,t)=\psi^0_\pm \left[1 + \varphi_\pm(x,t)\right]
\exp(-i\, \mu\, t/\hbar),
\end{equation}
where $|\varphi_\pm(x,t)| \ll 1$. In the absence of external potential
($U_{\rm ext}=0$), the $\varphi_\pm(x,t)$'s which are solutions
of the linearized version of Eq.~\eqref{eq:polar3} are plane waves
whose wavevector $q$ and frequency $\omega$ are related by
\begin{align}\label{polar7}
0=& \left. \omega^4+2\,i\,\gamma\, \omega^3
- \left(\frac{q^4}{2}+\frac{2}{1+\alpha}\, q^2\right)\omega^2 \right. \notag
\\
&
-2\,i\, \gamma \left(\frac{q^4}{4}+
4 \, \varrho_+^0 \, \varrho_-^0 \,
\frac{1-\alpha}{1+\alpha}\, q^2\right) \omega
\nonumber \\ &
+
\frac{q^4}{4}\left(\frac{q^4}{4}+\frac{2}{1+\alpha}\,q^2
+16\, \varrho_+^0\,\varrho_-^0 \, \frac{1-\alpha}{1+\alpha}
\right).
\end{align}
In this equation we note $\alpha=\alpha_2/\alpha_1$ ($-1<\alpha<0$),
$\varrho^0_\pm = \rho^0_\pm/\rho^0 =\frac{1}{2} (1 \pm
\Omega/\Omega_{\rm crit})$, and we use dimensionless quantities:
Energies are henceforth expressed in units of $\mu$, lengths in units
of $\xi$ [where $\xi=\hbar/(m\, \mu)^{1/2}$], and velocities in units
of $(\mu/m)^{1/2}$. Equation \eqref{polar7} has already been obtained
in Ref.~\onlinecite{Fil05} in the case of a two-component Bose gas
(i.e., in the absence of damping: $\gamma=0$) without magnetic
field.

\begin{figure}[h!]
\includegraphics*[width=0.99\linewidth]{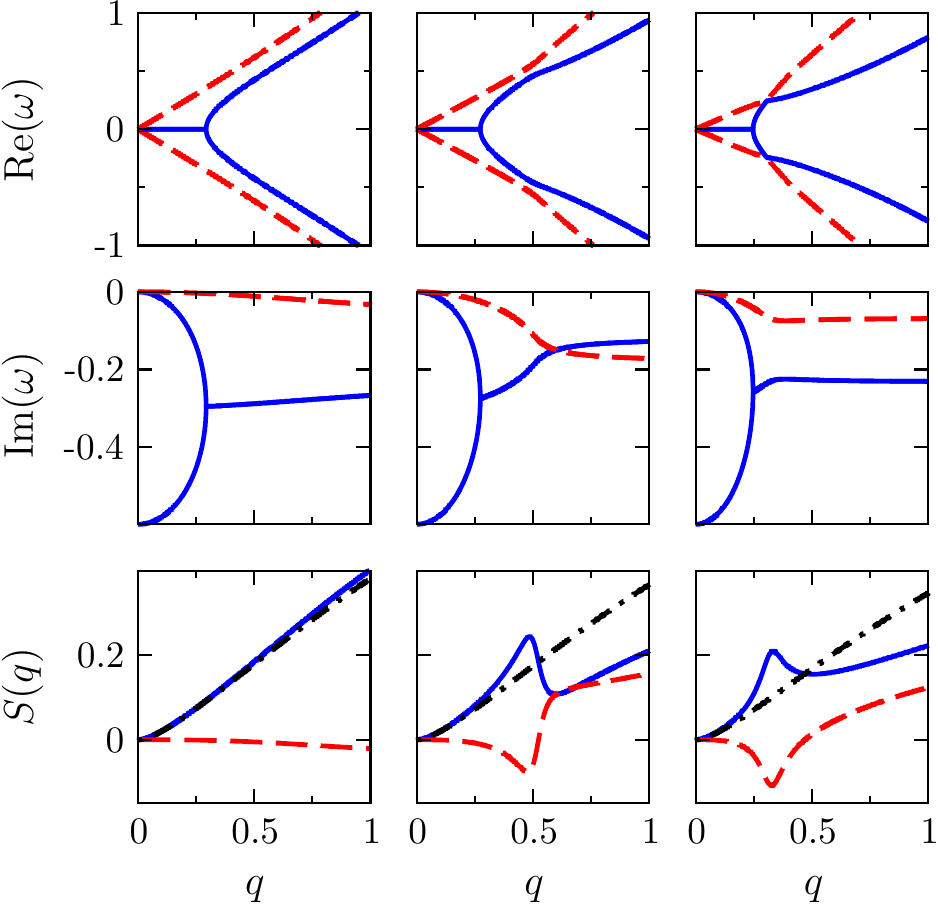}
\caption{(Color online) Dispersion relations in the case $\alpha=-0.2$
  and $\gamma=0.3$ \cite{remark1}. The first (second) row displays
  $\mbox{Re}(\omega)$ [$\mbox{Im}(\omega)$] as a function of $q$. The
  lower row displays the contribution of the density and polarization
  modes to the total density structure factor $S(q)$ (black dot-dashed
  line). In each plot the blue solid curve corresponds to the density mode
  and the red dashed curve to the polarization mode. The three columns
  are drawn in the cases (from left to right) $\Omega/\Omega_{\rm
    crit}=0.2$, 0.5, and 0.7.} 
\label{fig_disp}
\end{figure}

Solving the fourth-degree equation \eqref{polar7} yields the
dispersion relations $\omega=\omega(q)$. If $\omega(q)$ is a solution,
then $-\omega^*(q)$ is also a solution. As a result, the solutions
come into pairs having either the same zero real part or the same
imaginary part and opposite real parts. Some typical dispersion
relations are plotted in Fig.~\ref{fig_disp}.  In the limit of weak
magnetic field one pair of solutions corresponds to the usual
density-fluctuation mode (in which both components oscillate in
phase), the other one to a polarization-fluctuation mode (with
counterphase oscillations of the two components). 
We henceforth keep using the denominations ``density mode'' and
``polarization mode'' although the separation between the two types of
fluctuations is less strict for finite magnetic field, as illustrated
in the lower row of Fig.~\ref{fig_disp} where we plot the contribution
of each mode to the static structure factor $S(q)=\int
S(q,\omega) \, d\omega$, where $S(q,\omega)$ is the (zero
temperature) dynamical structure factor \cite{Pit03}. The fact that
one of the contributions can be negative originates in the
non conservative nature of Eq.~(\ref{eq:polar3}), but it is
interesting to note that, despite its losses, the system keeps a
constant density and still verifies the $f$-sum rule: $\int_0^\infty
\omega \, S(q,\omega) \, d\omega=\rho_0 \, q^2/2$.

From Eq.~\eqref{polar7} one can show that the polarization mode is
undamped (i) when $\varrho_+^0 = \varrho_-^0 = \frac{1}{2}$, i.e., in
the absence of magnetic field, and (ii) when $\varrho_-^0=0$, i.e., at
the critical magnetic field.  This is a first hint indicating that the
damping of the polarization mode is weak.  One can further show that
this damping is zero up to order $O(\Omega^2)$ in the external
magnetic field. A final evidence comes from the fact that the damping
of the polarization mode is always zero in the long-wavelength limit
as we discuss now.  In the absence of damping and of magnetic field
($\gamma=0$ and $\Omega=0$, respectively) the long-wavelength behavior
of both modes corresponds to a linear dispersion relation: The system
exhibits two types of sound.  One is the usual sound of velocity
$c_{\rm d}=1$. The other is the ``polarization sound'' of velocity
$c_{\rm p} = [(1-\alpha)/(1+\alpha)]^{1/2}$. For nonzero $\gamma$ the
usual sound waves are damped; this is not the case for the
polarization sound, as clearly seen in Fig.~\ref{fig_disp}. In
the general case where $\gamma$ and $\Omega$ are nonzero, looking for
a solution of Eq.~\eqref{polar7} under the form $\omega(q)=c_{\rm
  p}\,q$, in the limit $|q|\ll 1$ and $c_{\rm p} |q| \ll \gamma$ one
gets $c_{\rm p} = [(1-\Omega^2/\Omega_{\rm crit}^2)
(1-\alpha)/(1+\alpha)]^{1/2}$.

\begin{figure}
\includegraphics*[width=0.99\linewidth]{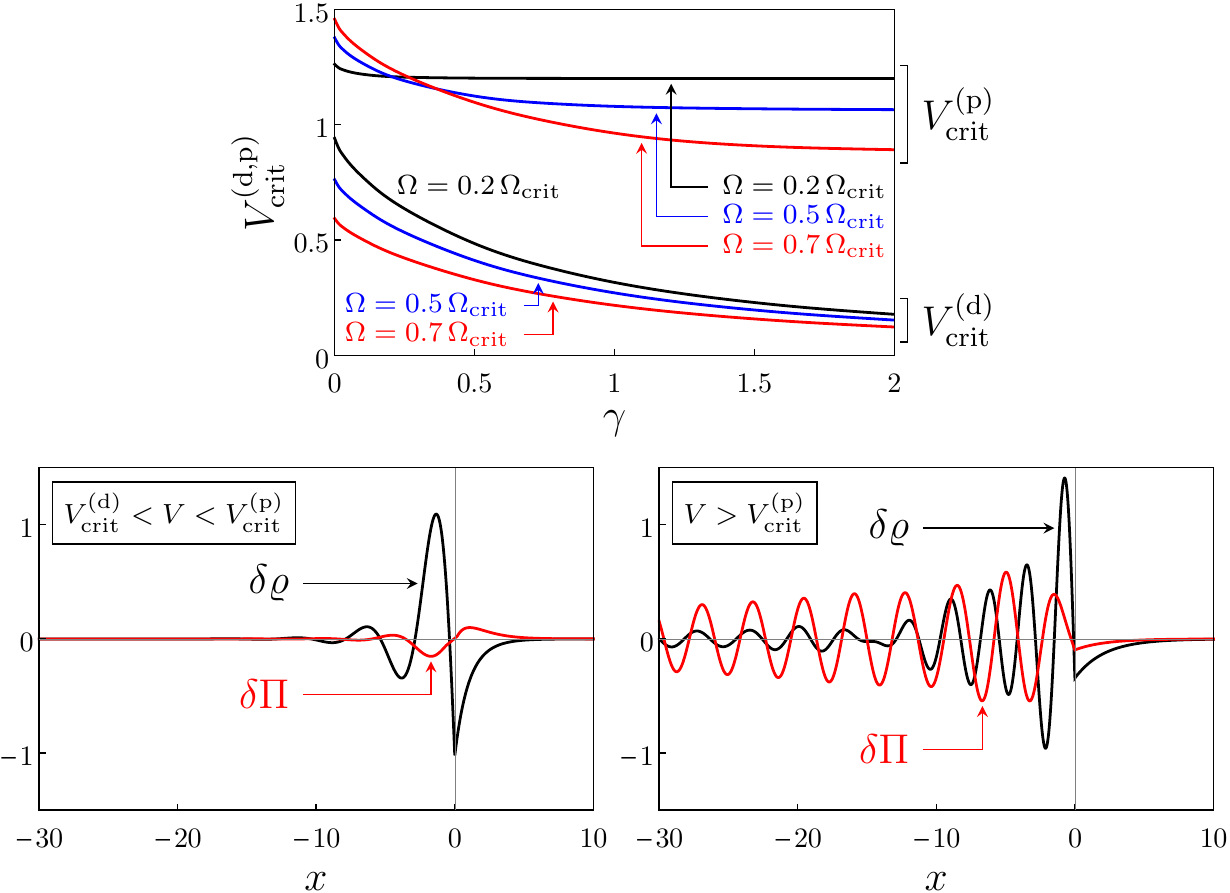}
\caption{(Color online) Upper panel: Critical velocities $V_{\rm
    crit}^{({\rm d})}$ and $V_{\rm crit}^{({\rm p})}$ as a function of
  $\gamma$ for different strengths of the magnetic field. The plot
  is drawn when $\alpha=-0.2$. Lower panels: Rescaled fluctuations of
  the density [$\delta\varrho(x)/\varkappa$: black curves] and of the
  polarization [$\delta\Pi(x)/\varkappa$: red curves] induced by a
  $\delta$-peak potential $\varkappa\, \delta(x+Vt)$. The modulation
  patterns are drawn for $\alpha=-0.2$, $\gamma=0.2$, and
  $\Omega/\Omega_{\rm crit}=0.1$. In this case, $V_{\rm crit}^{({\rm
      d})}=0.69(4)$ and $V_{\rm crit}^{({\rm p})}=1.21(9)$. The
  figures are drawn in the frame where the obstacle stays at rest at
  the origin and where the polariton fluid moves from left to right at
  velocity $V>0$. For the
  lower left panel $V=1.1$ and for the lower right panel $V=1.5$.}
\label{crit-V}
\end{figure}

From the knowledge of the dispersion relations one can compute the
linear response function $\chi_{\pm}(q,\omega)$ which characterizes
how the rescaled density $\varrho_{\pm}(x,t)=\rho_{\pm}(x,t)/\rho^{0}$
responds to a weak external scalar potential with wavevector $q$ and
pulsation $\omega$. This makes it possible to determine the wake
generated by a weakly perturbing obstacle moving at constant velocity
$V$ with respect to the polariton fluid. We do not detail the
computation which has been presented in Ref.~\onlinecite{Lar12} in the
case of a scalar order parameter.  In the present case, there exist
two particular velocities corresponding to the opening of channels of
(damped) Cherenkov radiation: $V_{\rm crit}^{({\rm d})}$ is the
threshold for emission of density waves and $V_{\rm crit}^{({\rm p})}$
is the threshold for emission of polarization waves.  These velocities
are functions of the losses in the system (i.e., of $\gamma$) and of
the strength of the external magnetic field (i.e., of $\Omega$). They
are represented in Fig.~\ref{crit-V}.

The physical meaning of these velocities 
 can be verified by inspecting the perturbations induced
by the obstacle which are represented in Fig.~\ref{crit-V} in the
simplest case where the external potential is of the form $U_{\rm
  ext}=\varkappa\, \delta(x+Vt)$. The plots are drawn in the frame
where the obstacle is at rest at the origin and where the polariton
fluid moves from left to right at velocity $V>0$. In this frame the
perturbations are stationary. In Fig.~\ref{crit-V} we do not
display separately $\delta\varrho_{+}=\varrho_{+}-\varrho_+^{0}$ and 
$\delta\varrho_{-}=\varrho_{-}-\varrho_-^{0}$ but we
rather plot the relevant physical observables: the fluctuations of the
total density ($\delta\varrho=\delta\varrho_{+} + \delta\varrho_{-}$)
and of the polarization ($\delta\Pi = \delta\varrho_{+} -
\delta\varrho_{-}$). The critical velocity $V_{\rm crit}^{({\rm p})}$
being larger than $V_{\rm crit}^{({\rm d})}$, an obstacle whose
velocity $V$ relative to the condensate lies between these two
critical velocities (such as considered in the lower left plot of
Fig.~\ref{crit-V}) only emits density fluctuations \cite{remark2}. 
On the contrary,
when $V$ is larger than both $V_{\rm crit}^{({\rm d})}$ and $V_{\rm
  crit}^{({\rm p})}$, the wake consists in both density and
polarization fluctuations (see Fig.~\ref{crit-V}, lower right
plot). We also note that a direct computation of the density patterns
$\varrho_{\pm}$ for several intensities of the magnetic field shows
that, as stated above, the polarization wave is weakly damped at low
and at high field, facilitating the experimental observation of the
polarization signal compared to that of density fluctuations.

\begin{figure}
\includegraphics*[width=0.99\linewidth]{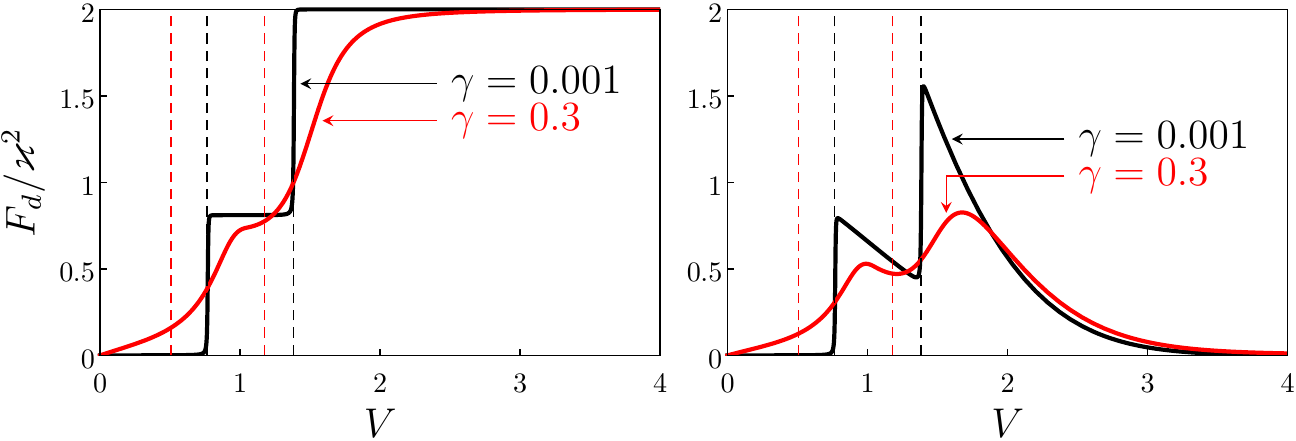}
\caption{(Color online) Drag force $F_d/\varkappa^2$ as a function of
  the velocity $V$ of the obstacle relative to the condensate for two
  damping parameters: $\gamma=0.001$ (solid black curves) and
  $\gamma=0.3$ (solid red curves). The left plot corresponds to a
  point-like obstacle and the right plot to a Gaussian potential of
  width $\ell=0.5$. The computation is done for $\alpha=-0.2$ and
  $\Omega=0.5\, \Omega_{\rm crit}$. In this case $V_{\rm crit}^{({\rm
      d})}=0.76(2)$ and $V_{\rm crit}^{({\rm p})}=1.37(7)$ when
  $\gamma=0.001$, whereas $V_{\rm crit}^{({\rm d})}=0.50(5)$ and
  $V_{\rm crit}^{({\rm p})}=1.17(3)$ when $\gamma=0.3$. All these
  threshold velocities are indicated by vertical colored dashed lines
  in the figure.}
\label{Drag}
\end{figure}

The existence of two critical velocities has also an important effect
on the behavior of the drag force $F_d$ experienced by the
obstacle. This is illustrated in Fig.~\ref{Drag} where $F_d$ is
plotted as a function of $V$ for two types of obstacles: a point-like
scatterer of intensity $\varkappa$, for which $U_{\rm
  ext}=\varkappa\,\delta(x+Vt)$, and a Gaussian potential of same
intensity and of width $\ell$, for which $U_{\rm
  ext}=\frac{\varkappa}{\ell\sqrt{\pi}} \exp[-(x+Vt)^2/\ell^2]$.
One sees in Fig.~\ref{Drag} that,
at very weak damping, $F_d$ is negligible at small velocity and shows
pronounced thresholds when $V$ reaches the critical velocities $V_{\rm
  crit}^{({\rm d})}$ and $V_{\rm crit}^{({\rm p})}$, demonstrating
that in the limit $\gamma\to 0$ the drag uniquely consists in wave
resistance. This corresponds to the Landau criterion for the onset of
dissipation: At each opening of a radiation channel (i.e., at $V=V_{\rm
  crit}^{({\rm d})}$ and $V_{\rm crit}^{({\rm p})}$) the drag is
abruptly increased. This reflects the work imparted to the fluid and
dissipated by generating the wave pattern which irreversibly radiates
energy away from the obstacle. For finite values of $\gamma$ instead,
the flow is never truly superfluid: The obstacle experiences a finite
force even at low velocity \cite{Wou10,Ber12}, which corresponds to
diffusion of momentum, i.e., to a viscous drag. In this case there is
no Landau criterion, but the system exhibits a smooth crossover from a
drag dominated by viscous-like phenomena (at low velocity) to one
dominated by wave resistance (at large velocity). Thus, it is more
appropriate to term the velocities $V_{\rm crit}^{({\rm d})}$ and
$V_{\rm crit}^{({\rm p})}$ Cherenkov- (or Mach-) rather than
Landau-critical velocities.

We also note that in the absence of external magnetic field (in the
case $\Omega=0$, not shown in the figure) no
step is seen in the drag, even for $\gamma\to 0$. This is due to the
fact that, despite the opening of a new radiation channel at $V
= V_{\rm crit}^{({\rm d})}$, the external scalar potential
cannot excite polarization waves in this case for symmetry reasons,
since no term in Eq.~\eqref{eq:polar3} can distinguish the spin-up
from the spin-down component when $\Omega=0$. This is reminiscent of
what occurs for the first and second sound in superfluid HeII: The
second sound which corresponds to a temperature (and entropy) wave
cannot be excited by oscillations of the container wall, contrarily to
the usual density waves associated with the first sound; see, e.g.,
Ref.~\onlinecite{ll-6}.

Finally we emphasize that another effect of the existence of the spin
degree of freedom is revealed in the absence of obstacle by the
quantum fluctuations of the polarization. One can show that in a
homogeneous condensate in the absence of damping and of magnetic
field, $g^{(2)}_{\rm p}(x,x')=\langle:\!\!\delta\hat\Pi(x)\,
\delta\hat\Pi(x')\!\!  :\rangle$ is a universal function of $c_{\rm
  p}|x-x'|$ which goes to zero when $|x-x'|\to\infty$. One gets $
g^{(2)}_{\rm p}(x,x)= -\frac{2}{\pi}\,c_{\rm p}<0$: This corresponds
to local sub-Poissonian fluctuations of the polarization. These
fluctuations are strongly modified in the presence of a (polarization)
sonic horizon. They acquire nonlocal features associated to the
correlated emission of analogous Hawking radiation, as first shown in
Refs.~\onlinecite{correlations} for density-density correlations.  The
present results suggest that in polariton systems the
polarization-polarization correlation function $g^{(2)}_{\rm p}(x,x')$
should be a quite efficient observable for witnessing Hawking
radiation, even in the absence of an external magnetic field \cite{Lar13}.

\begin{acknowledgments}
  We thank A.~Amo, J.~Bloch and S.~Stringari for fruitful discussions.
  This work was supported by the French ANR under grant n$^\circ$
  ANR-11-IDEX-0003-02 (Inter-Labex grant QEAGE).
\end{acknowledgments}


\begin{thebibliography}{99}
\bibitem{remark1} This is a standard value for the dimensionless
  parameter $\gamma$ which is typically of order $\hbar/(\tau\mu)$,
  $\tau$ being the lifetime of a polariton ($\tau\sim 10$ ps) and
  $\mu$ the chemical potential ($\mu\sim 0.5$ meV).
\bibitem{remark2} This is not quite true because, as explained above,
  the distinction between density and polarization modes is not sharp.
\end{thebibliography}
\end{document}